\documentstyle[prl,aps,epsfig,floats]{revtex}
\def\DESepsf(#1 width #2){\epsfxsize=#2 \epsfbox{#1}}
\begin{document}
\renewcommand{\thefootnote}{\fnsymbol{footnote}}

\pagestyle{empty}                                      
\draft
\vfill
\twocolumn[\hsize\textwidth\columnwidth\hsize\csname 
@twocolumnfalse\endcsname
\title{Window on Higgs Boson: Fourth Generation $b^\prime$ Decays Revisited}
\vfill
\author{Abdesslam Arhrib
%\footnote{On leave of absence from Department of 
%Mathematics FSTT, P.O. Box 416, Tangier, Morocco} 
 and Wei-Shu Hou }
\address{
\rm Department of Physics, National Taiwan University,
Taipei, Taiwan 10764, R.O.C.
}

%\date{\today}
%
%\vskip -1cm
%
\vfill
\maketitle
\begin{abstract}
Direct and indirect searches of the Higgs boson suggest 
that 113 GeV $\lesssim m_H \lesssim$ 170 GeV is likely.
With the LEP era over and the Tevatron Run II search via
$p\bar p \to WH+X$ arduous,
we revisit a case where $WH$ or $ZH +$ jets could arise via 
strong $b^\prime\bar b^\prime$ pair production. 
In contrast to 10 years ago,
the tight electroweak constraint on $t^\prime$--$b^\prime$ 
(hence $t^\prime$--$t$) splitting 
reduces FCNC $b^\prime\to bZ$, $bH$ rates,
making $b^\prime\to cW$ naturally competitive.
Such a ``cocktail solution" is precisely the mix
that could evade the CDF search for $b^\prime\to bZ$,
and the $b^\prime$ may well be lurking below the top.
In light of the Higgs program,
this two-in-one strategy should be pursued.
%, starting with revisiting Run I data.
\end{abstract}
\vskip2pc]

\pagestyle{plain}

%%% Introduction

The search for the Standard Model (SM) Higgs boson
is the holy grail of present day high energy physics.
%, has become a high stakes contest. % of late.
%
%Before the LEP era, 
%experimental knowledge of the Higgs boson was scant.
%After 12 years of LEP, Tevatron and SLC running, 
%we have learned a lot about the Higgs boson. 
The global electroweak (EW) fit, assuming SM, 
gives \cite{osaka2} $m_H = 62^{+53}_{-39}$~GeV, 
or $m_H < 170$ GeV at $95\%$ confidence level (CL).
Together with the 
direct search limit~\cite{osaka} of $m_H > 113.3$~GeV at LEP,
%there is anticipation that 
the Higgs boson seems ``just around the corner".
Surely enough, 
just before LEP shutdown,
there were exciting hints~\cite{aleph,www} for $m_H \simeq 115$~GeV,
and the shutdown was postponed by a month. 
The extra data collected did not greatly strengthen the case, 
but it was argued that a further run with 
about 200 pb$^{-1}$ per experiment at $\sqrt{s} =$  208.2 GeV 
might lead to a 5$\sigma$ discovery~\cite{www}.
Unfortunately, the wish was not granted,
fearing it might jeopardize the schedule for LHC 
--- the main goal of which is the Higgs boson!
As LEP is now closed, 
we have to wait for Run II at Tevatron which starts in 2001,
or the turn on of LHC in 2005.

Not surprisingly,
Higgs boson search in the 110--190 GeV range is now 
one of the prime objectives for Tevatron Run~II.
The main process, $q\bar{q}\to V^*\to V H$ ($V=W$, $Z$)
followed by $H\to b\bar{b}$ and
leptonic $V$ decays \cite{stange},
suffers from a small electroweak cross section: 
$\sigma(q\bar{q}\to W^\pm H) \simeq$ 0.3 to 0.002 pb 
at Run II energies for 100 GeV $\leq m_H \leq$ 200~GeV.
%Thus, direct $WH$ or $ZH$ production did not seem promising
%against the QCD background for typical Higgs boson decay modes. 
%With the aim at circumventing these problems,
A very recent study \cite{tevatronH} claims that,
with suitable cuts that exploit the 
kinematic differences between signal vs background,  
a statistically significant signal can be extracted
if one has {\it sufficient} luminosity. 
However, with the 2~fb$^{-1}$ expected by end of 2002,
one can barely rule out the LEP hint of $m_H \sim 115$ GeV at 95\%~CL, 
while to have a $5\sigma$ discovery, one would need 10--15~fb$^{-1}$.
Since this seems to be the amount of data one may realistically expect
% (with or without a major upgrade) 
before the start of LHC physics,
the Tevatron path to Higgs search is % rather 
arduous.

It is clear that a premium should be placed on processes with higher
cross sections (but manageable background)
that could aid the Higgs search at the Tevatron.
%One in general faces the issue of signal vs background.
%For example, the higher cross section of $gg\to H$ fusion 
%is severely limited by QCD background in a hadronic environment.
In this Letter we revisit a case~\cite{hou1} where
the $WH$ or $ZH$ signatures arise from 
strong $p\bar{p}\to b^\prime\bar{b^\prime} +X$ production
of sequential fourth generation $b^\prime$ pairs, 
followed by $b^\prime \to cW$, $bZ$, $bH$ decays
that are all of comparable strength.
The effective $p\bar{p}\to WH +X$ or $ZH +X$ cross sections
could be at a few pb$^{-1}$.
%, hence considerably higher than $p\bar{p}\to VH + X$.  

The main {\it new} observations we make are as follows.
EW precision data give stringent constraint on 
$m_{t^\prime} - m_{b^\prime}$~\cite{PDG}.
For $m_{b^\prime} < m_t$,
the $t^\prime$--$t$ splitting is considerably smaller
than assumed in \cite{hou1},
and Glashow-Illiopoulos-Maiani (GIM) suppression
of $b^\prime\to bZ$, $bH$ decays is more severe\cite{PR},
hence the $b^\prime \to cW$ channel becomes more prominent.
Furthermore, recent direct search by CDF \cite{CDF} has ruled out
$b^\prime \to b Z$ decay for $m_{b^\prime} <$ 200 GeV 
if ${\cal B}(b^\prime\to bZ) =$ 100\%,
but may be evaded if $b^\prime \to cW$ (but {\it not} $b^\prime \to bH$)
dilutes ${\cal B}(b^\prime\to bZ)$.
Note that
a dominant but not predominant~\cite{D0} $b^\prime \to cW$ decay 
could help explain some irregularities of the $t\bar t$ signal. 
We argue that
the Cabibbo-Kobayashi-Maskawa (CKM) mixing element 
$V_{cb^\prime} \sim 10^{-3}$ is plausible,
and is just the right amount to allow
a ``cocktail solution" of $b^\prime\to c W$, $bZ$ and $bH$
that can evade the CDF bound.
This offers new possibilities for Higgs search 
up to $m_H < m_{b^\prime}-m_b$.

%%% The Cocktail Solution

It is known that, if the $b^\prime$ quark
exists and $m_{b^\prime} < m_t$, 
it may decay in unusual ways:
the charged current (CC) $b^\prime \to tW$ decay 
is kinematically forbidden, 
the $b^\prime \to cW$ decay is highly Cabibbo suppressed,
hence flavor changing neutral current (FCNC) 
$b^\prime \to b$ transitions would likely dominate~\cite{hou}.
The suggestion was pursued~\cite{PDG} by collider experiments at 
Tristan, SLC, LEP and Tevatron,
with LEP setting the unequivocal bound of $m_{F} > m_Z/2$ on
all new fermions $F$ that couple to the $Z$.
The D0 Collaboration \cite{D01} excluded the range 
$m_Z/2 < m_{b^\prime} < m_Z + m_b$ by a null search for
$b^\prime \to b \gamma$ and $b^\prime \to b g$.
%
%But is a $b^\prime$ lighter than the top still viable?
We remark that,
with $N_\nu \cong 3$ as measured by SLC and LEP since 1989, 
the existence of a sequential fourth generation is 
not strongly motivated (For a recent review, see \cite{PR}).
However, the observation of neutrino oscillations
does imply an enlarged neutrino sector.
A more important motivation comes from the intense
competition for Higgs search as stated above.

For $m_{b^\prime} > m_Z+ m_b$, the decay
$b^\prime \to b Z$~\cite{hou} is expected to 
dominate over the other FCNC decay processes,
except for $b^\prime \to b H$~\cite{hou1,ehs}
if $m_{b^\prime} > m_H+ m_b$ also.
Recently, the CDF Collaboration \cite{CDF}
gave an upper limit on the product 
$\sigma(p\bar{p}\to b^\prime\bar{b^\prime})
 \times [{\cal B}(b^\prime\to bZ)]^2$ 
as a function of $m_{b^\prime}$, 
which excludes at 95\% CL the range 100 GeV $ < m_{b^\prime} <$ 199 GeV
if ${\cal B}(b^\prime\rightarrow bZ) = 100\%$.
For ${\cal B}(b^\prime \to b H) \neq 0$,
so long that ${\cal B}(b^\prime \to b Z)$ does not vanish,
the CDF bound still largely applies
since hadronic final states of $b^\prime \to b Z$ and $bH$ 
are rather similar, and in fact the $bH$ mode
has better $b$-tagging efficiency.
What CDF apparently did not pursue in any detail
is the $b^\prime \to cW$ possibility.
Clearly the $b$-tagging efficiency for $cW$ mode 
would be much worse than $bZ$ or $bH$.
Since $b$-tagging is an important part of the
CDF $b^\prime \to b Z$ search strategy,
{\it one may evade the CDF search if ${\cal B}(b^\prime \to cW)$ is sizable}.

Precision EW data provide stringent constraints on
the fourth generation:
there is a $2.5\sigma$ discrepancy 
between $S=-0.07\pm 0.11$~\cite{PDG}
and $ S=2/3\pi \cong 0.21$ for a heavy degenerate fourth generation.
However, using exact expressions for 
gauge boson self-energies for 
$m_{b^\prime}=m_{t^\prime}=150$ GeV, 
$m_{E}=200$ and $m_{N}=100$~GeV 
($E$, $N$ are fourth generation charged and neutral leptons), 
one finds \cite{PR} $S \approx 0.11$ instead of 0.21, 
and the discrepancy drops below $2\sigma$.
Given the excellent agreement between SM and EW data, 
a discrepancy at this level
in a few measurables is not tantalizing.
For $S=0.2$, the  $2\sigma$ upper bound on $T$ is 
approximately 0.2 ($\delta\rho\approx$ 0.0015). 
Using the analytic expression
of $\delta\rho$ \cite{PDG},  %,peskin},  %erler}, 
we find 
$\Delta_Q=|m_{t^\prime}-m_{b^\prime}| \leq 60$~GeV and 
$\Delta_L=|m_{E}-m_{N}| \leq 104$~GeV,
which can be weakened 
if we just take the $\delta\rho$ constraint.
At the $2\sigma$ level and for $m_H \lesssim 1$ TeV, 
one finds~\cite{PDG} %,erler}  
$\rho_0=0.9998_{-0.0012}^{+0.0034}$ or $-0.0014 < \delta\rho< 0.0032$, 
hence $\Delta_Q \leq 86$ GeV and $\Delta_L \leq 148$ GeV.

In the following, we shall take the conservative range
$\Delta_Q=|m_{t^\prime}-m_{b^\prime}| \leq 60$~GeV.
For $m_{b^\prime} < m_t$,
% the scenario $m_t \sim m_{b^\prime} \sim m_{t^\prime}$ is realized, and 
this implies that the {\it $t^\prime$--$t$ splitting is 
far less \cite{PR} than assumed 10 years ago} \cite{hou1},
and FCNC $b^\prime \to bZ$, $bH$ decays are more GIM suppressed.
Thus, the CC $b^\prime\to cW$ mode, 
though highly Cabibbo suppressed, can be more competitive.
Note that $m_t \sim m_{b^\prime} \sim m_{t^\prime}$
would {\it in general imply near maximal mixing},
or $V_{tb^\prime} \simeq V_{t^\prime b} \simeq 1/\sqrt{2}$.

What is the ``natural" strength of $V_{cb^\prime}$?
We cannot know for certain,
but give two plausible arguments here:
Since each involve two generation jump,
perhaps $V_{cb^\prime} \sim V_{ub}$;
or one could guess that 
$V_{cb^\prime} \sim m_s/m_{b^\prime}$ 
since $V_{cb} \sim m_s/m_b$ \cite{babu}.
Both cases suggest $V_{cb^\prime} \sim 10^{-3}$,
just what is needed to make $b^\prime \to cW \sim b^\prime \to b Z$
in rate, as we will show. 
Thus,
EW precision data, while not strongly supporting the
existence of a fourth generation,
together with the hierarchical pattern of quark masses and mixings,
{\it lead naturally to the ``cocktail solution"}
of $b^\prime \to cW \sim bZ \gtrsim bH$
that can evade CDF search for $b^\prime \to bZ$.
This is in contrast to previous expectations \cite{hou1,hou,ehs} 
that $b^\prime \to cW$ would be considerably 
below $b^\prime \to  bZ,\ bH$ decays.
We remark here that
comparison of $b^\prime \to cW$ and $b^\prime \to  bZ$
were made recently in \cite{PR}, 
but not in conjunction with $b^\prime \to  bH$;
the importance of the latter mode was emphasized in \cite{sher}
in the context of evading CDF bound on $b^\prime \to  bZ$,
but a detailed discussion of
the ``cocktail solution" was not given.

%% Calculation

We perform a one-loop calculation of $b^\prime\rightarrow bZ$,
$bH$  using the FeynArts and FeynCalc \cite{seep} packages,
where the full set of diagrams (see Fig. 1 of \cite{hou1})
are generated and computed, 
and we use the FF--package~\cite{ff} for our numerical analysis.
We keep both external and internal masses, 
except the $m_b$ (and $m_c$, $m_u$) which 
can be neglected to good approximation.
Several analytic and numerical checks are carried out 
following \cite{hou1,hou,ehs}, with complete agreement found. 
Further numerical checks against
$t\to c H$, $c Z$ \cite{bar,Eilam} in SM again give full agreement. 
In the following, 
we take 115 GeV $< m_H < 170$ GeV,
$m_t =$ 175 GeV, $\Delta_Q = m_{t^\prime} - m_{b^\prime} \leq 60$ GeV, 
and $r_{\rm CKM} \equiv \vert V_{cb^\prime}/V_{tb}V_{tb^\prime}\vert$
in the $10^{-3}$ range.

%% Results

\begin{figure}[t!]
\smallskip
\centerline{%\hskip-0.8cm
            {\epsfxsize2. in  \epsffile{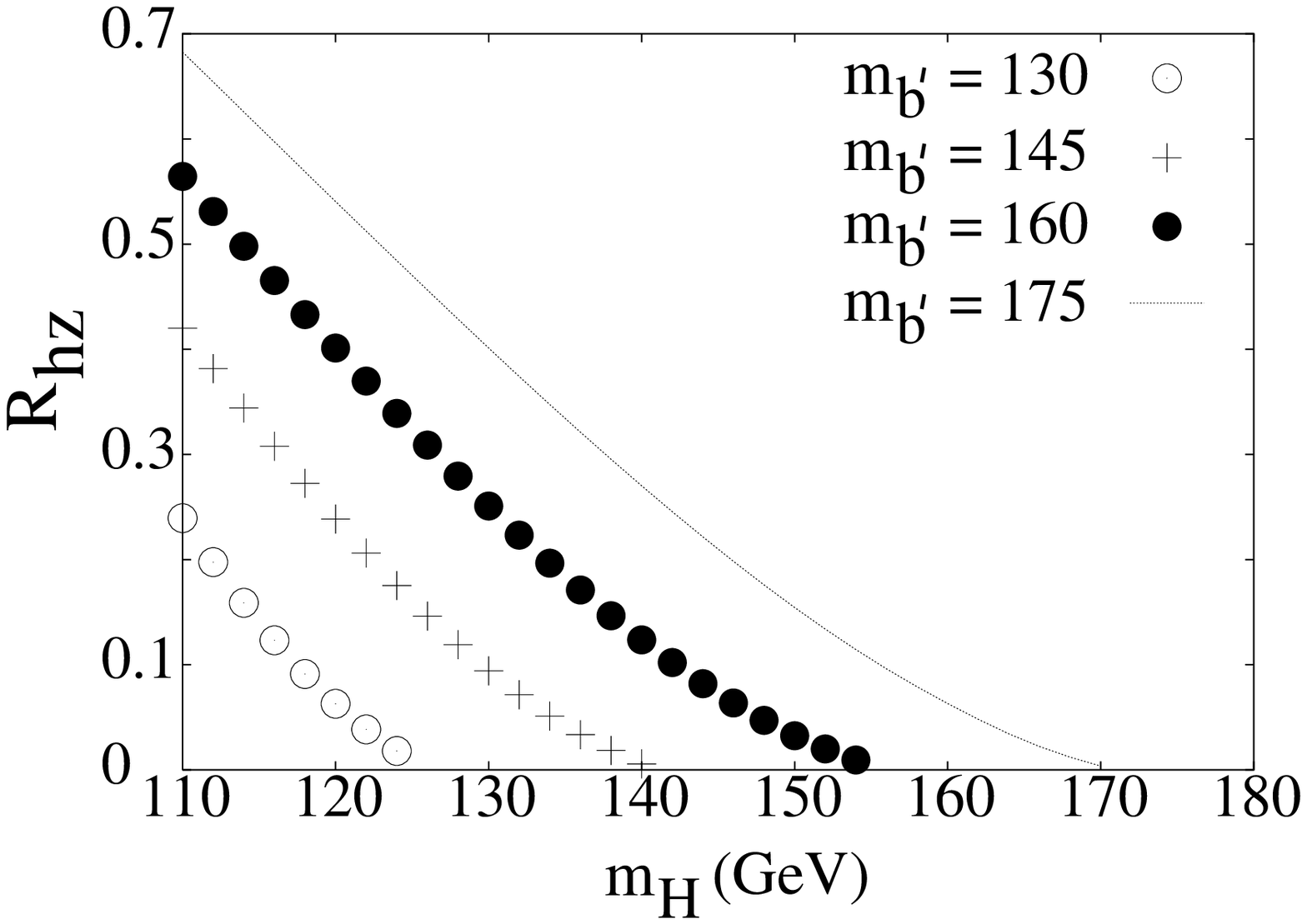}}}
\smallskip\smallskip\smallskip\smallskip\smallskip\smallskip\smallskip
\caption{
The ratio $R_{hz}=
\Gamma(b^\prime\rightarrow bH)/\Gamma(b^\prime\rightarrow bZ)$
vs $m_{H}$ for several $m_{b^\prime}$ values
and $\Delta_Q = m_{t^\prime} - m_{b^\prime} = 55$ GeV.
}
\end{figure}
%%%%%%%%%%%%%%

The CKM factors $\vert V_{t^\prime b}V_{tb^\prime}\vert
 \approx \vert V_{tb}V_{tb^\prime}\vert$ 
actually cancel in the ratio
%\begin{eqnarray}
$
R_{hz}=
\Gamma(b^\prime\rightarrow bH)/\Gamma(b^\prime\rightarrow bZ),
$
%\end{eqnarray}
and it depends on $m_{b^\prime}$, $m_{t^\prime}$ and $m_H$.
%with $m_{t^\prime}- m_{b^\prime}$ constrained by EW data.
In Fig. 1 we show $R_{hz}$ vs $m_H$
for several $m_{b^\prime}$ values
with $\Delta_Q$ fixed at 55 GeV.  
It is clear that, for light $m_H$ and 
relatively large $m_{b^\prime}$, 
$R_{hz}$ can be of order $0.5 - 1$,
which means $b^\prime\to bH$ is 
competitive with $b^\prime\to b Z$
so long that it is phase space allowed \cite{hou1}.

\begin{figure}[b!]
\centerline{\hskip0.4cm
\smallskip
            {\epsfxsize1.4 in \epsffile{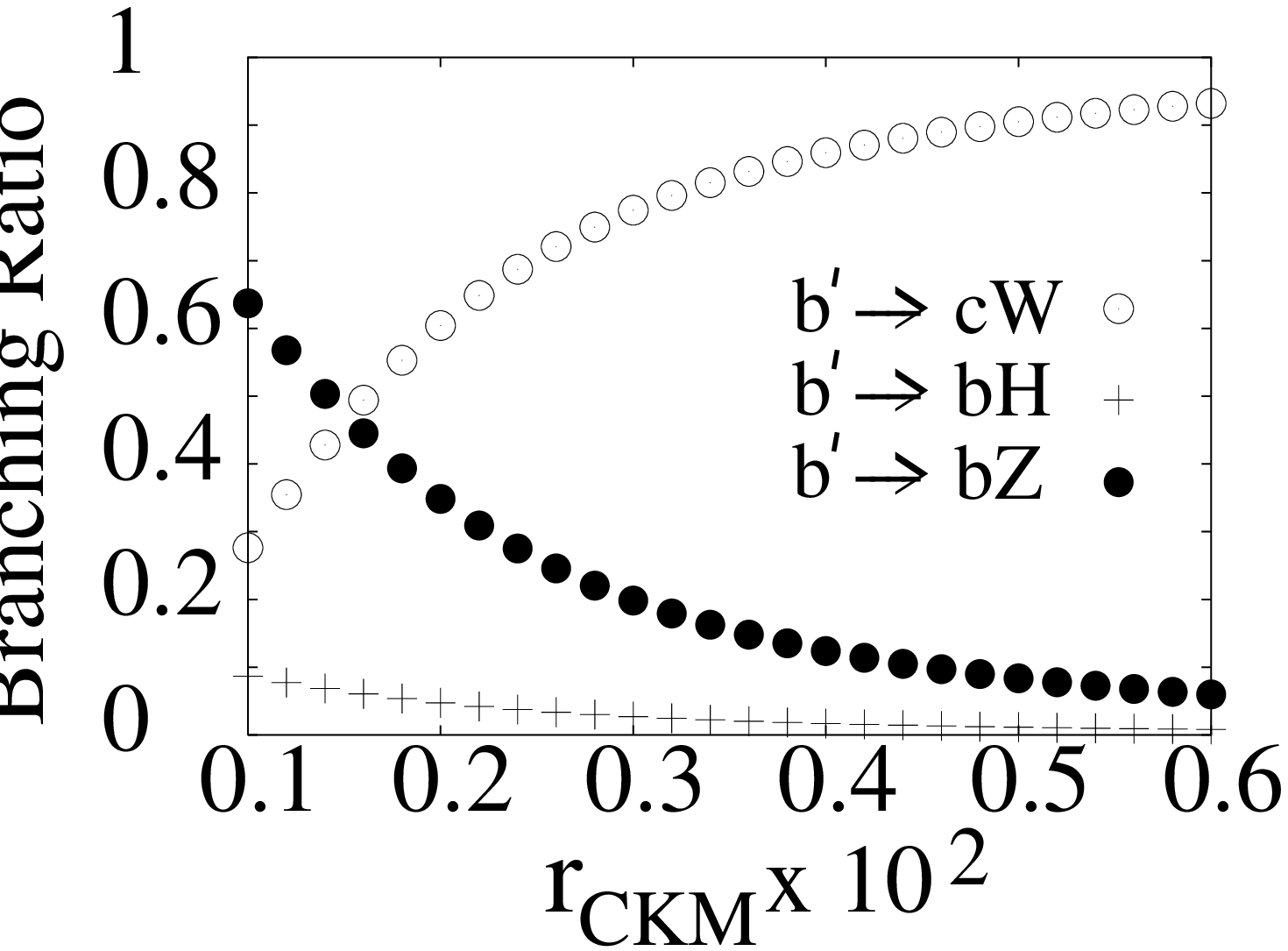}} \hskip0.4cm
            {\epsfxsize1.4 in \epsffile{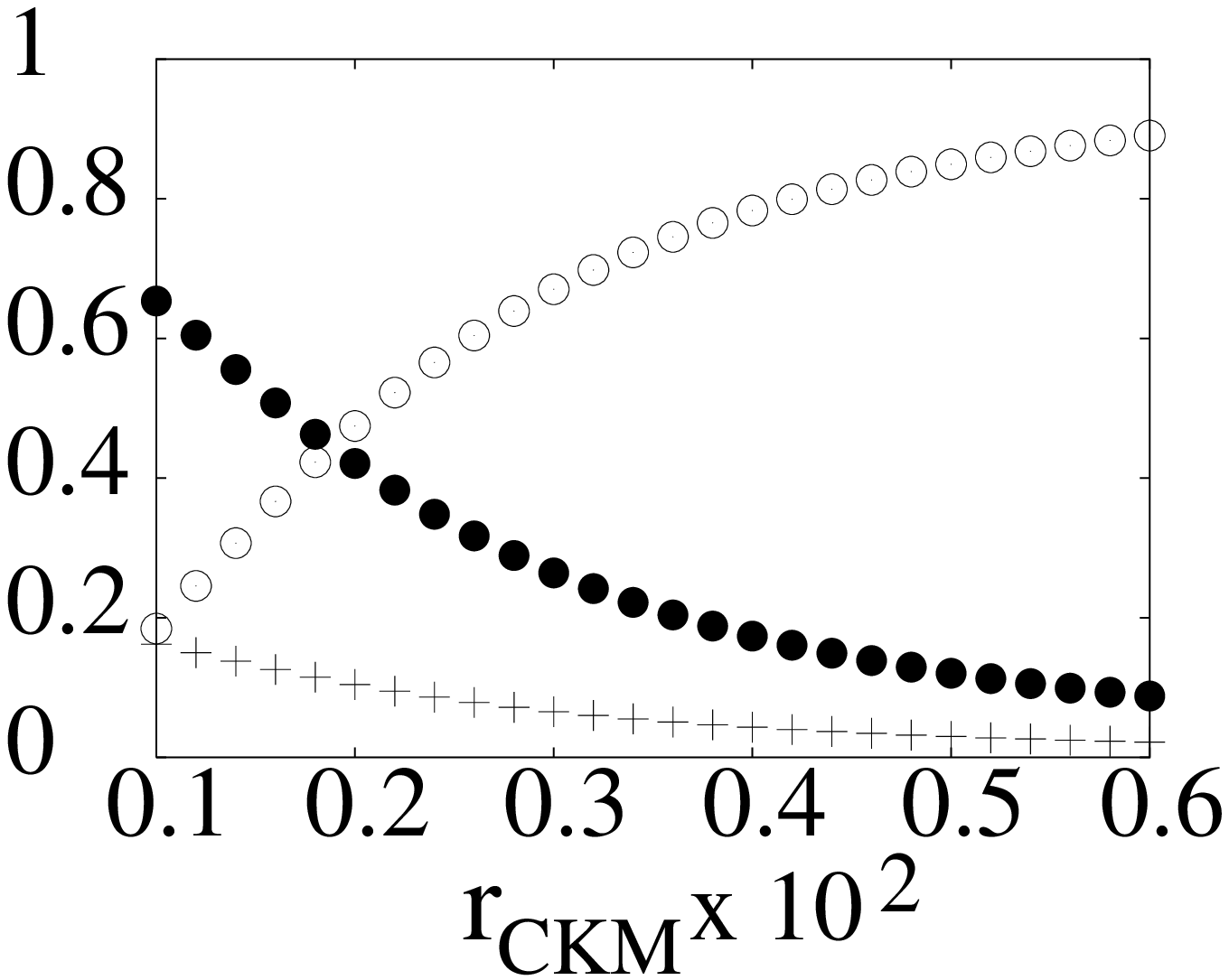}}}
\smallskip\smallskip\smallskip\smallskip\smallskip
\smallskip\smallskip
\caption{
${\cal B}(b^\prime \to \{cW,bZ,bH\})$ vs
$r_{\rm CKM}$ % = \vert V_{cb^\prime}/V_{tb}V_{tb^\prime}\vert$
for $m_{b'}$, $m_H$, $\Delta_Q =$ 130, 115, 20 GeV (left) and
160, 130, 40 GeV (right).
}
\end{figure}
%%%%%%%%%%%%%

For the actual branching ratios \cite{tW*},
\[
 {\cal B}(b^\prime\to \{cW,bZ,bH\}) =
 \Gamma(b^\prime\rightarrow \{cW, bZ, bH\} )/\Gamma_{b^\prime}, 
\]
we assume
$\Gamma_{b^\prime} = \Gamma(b^\prime\rightarrow cW) +
\Gamma(b^\prime\rightarrow bZ)+ \Gamma(b^\prime\rightarrow bH)$.
Since 
$\Gamma(b^\prime\rightarrow cW)        \propto |V_{cb^\prime}|^2$ while 
$\Gamma(b^\prime\rightarrow \{bZ,bH\}) \propto |V_{tb}V_{tb^\prime}|^2$,
the branching fractions depend critically on $r_{\rm CKM}$.  
In Fig. 2 we illustrate ${\cal B}(b^\prime \to \{cW,bZ,bH\})$ vs
$r_{\rm CKM} \equiv \vert V_{cb^\prime}/V_{tb}V_{tb^\prime} \vert$
for $m_{b^\prime}$, $m_H$, $\Delta_Q=$ 130, 115, 20 GeV 
and 160, 130, 40 GeV.
For $r_{\rm CKM} > 3\times 10^{-3}$, 
${\cal B}(b^\prime\to cW) > 80\%$
which would unlikely survive the standard top search.  
On the other hand, for 
$r_{\rm CKM}  \approx (1$--$3)\times 10^{-3}$, 
we find that  
${\cal B}(b^\prime\to bZ) \sim$  64\%--20\% (65\%--26\%), 
${\cal B}(b^\prime\to cW) \sim$   28\%--78\% (18\%--67\%), and 
${\cal B}(b^\prime\to bH) \sim$   8.7\%--2.7\% (16\%--6.5\%),
hence $b^\prime\to bZ$ and $cW$ are comparable,
and may allow the $b^\prime$ quark to 
evade the CDF $b^\prime \to bZ$ search.
%, as we will illustrate later.

\begin{figure}[t!]
\vskip0.15cm
\centerline{\hskip0.4cm
           {\epsfxsize1.4 in \epsffile{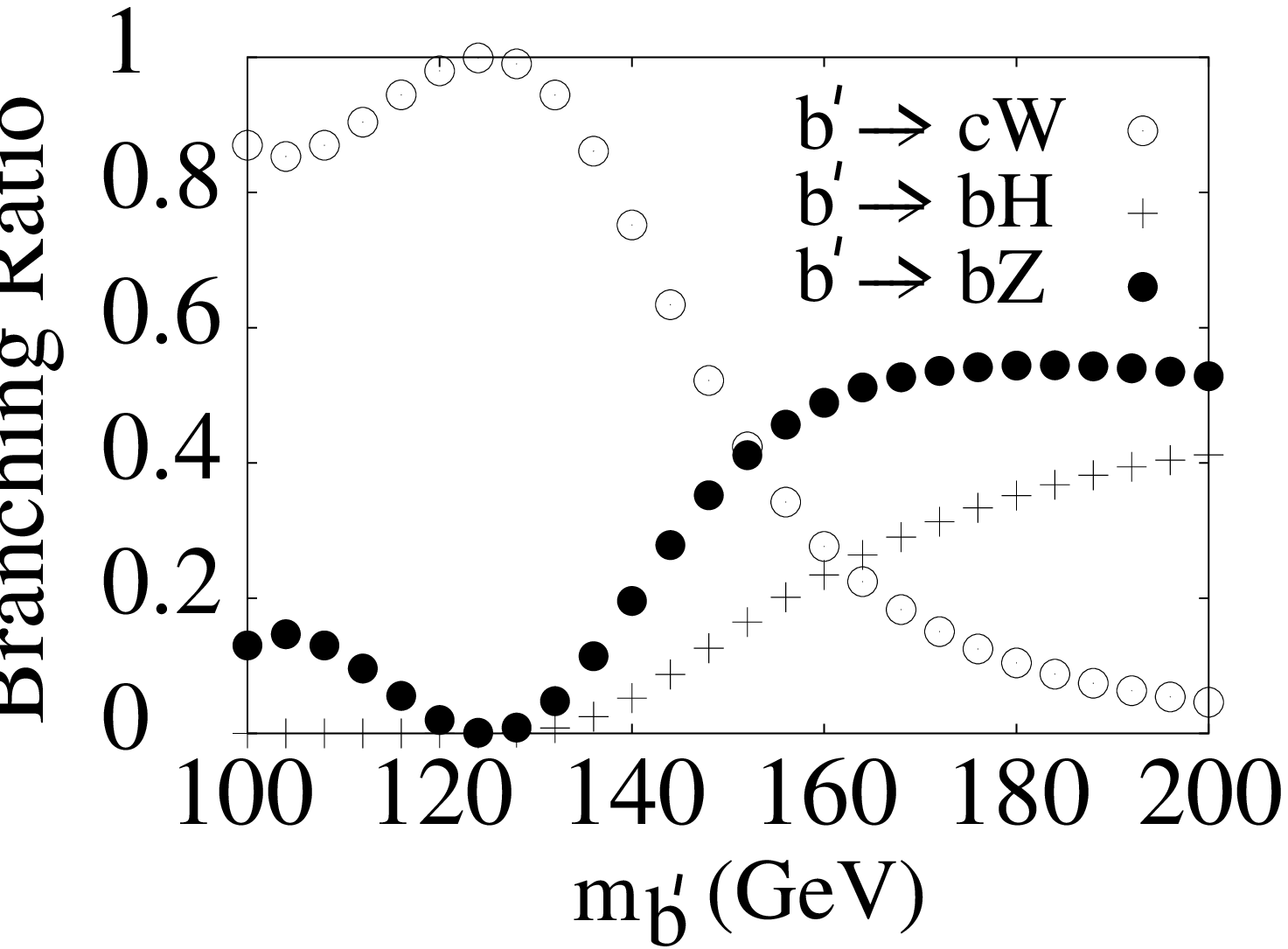}} \hskip0.4cm
            {\epsfxsize1.4 in \epsffile{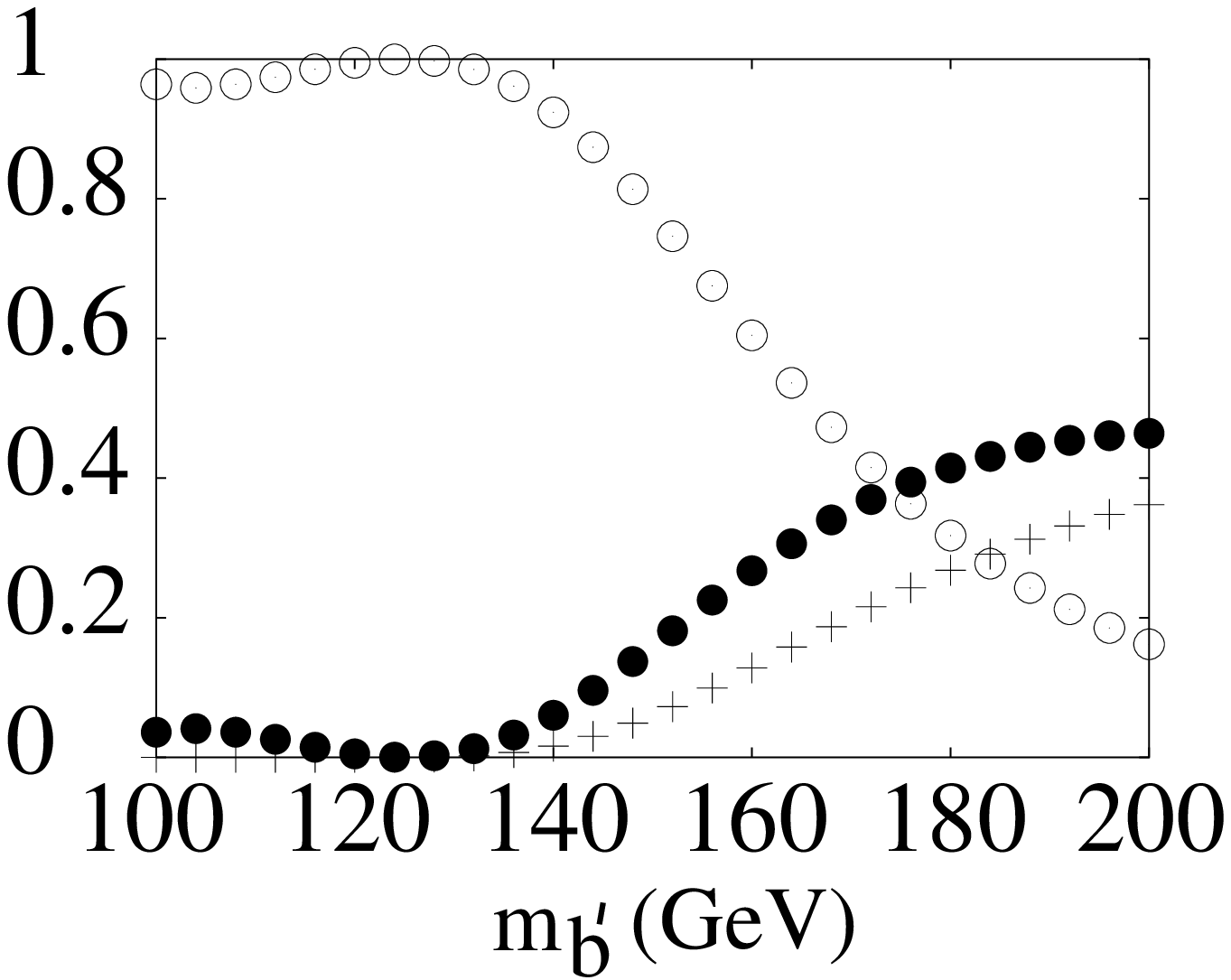}}}
\smallskip\smallskip\smallskip\smallskip\smallskip
\smallskip\smallskip
\caption{
${\cal B}(b^\prime \to \{cW,bZ,bH\})$ vs.
$m_{b^\prime}$ with $m_H,\ \Delta_Q=$ 115, 50 GeV
for $r_{\rm CKM}=$ 0.002 (left),  0.004 (right).
}
\end{figure}
%%%%%%%%%%%%%

We illustrate, in Fig. 3, 
${\cal B}(b^\prime \to \{cW,bZ,bH\})$ vs $m_{b^\prime}$ 
for two values of $r_{\rm CKM}$ with $\Delta_Q=50$ GeV held fixed. 
For 100 GeV $< m_{b^\prime} < 135$ GeV  hence 
150 GeV $< m_{t^\prime} < 185$ GeV,
$b^\prime\to bZ,\ bH$ are suppressed
by small $t^\prime$-$t$ splitting,
and $b^\prime\to cW$ predominance
can be seen from left-hand side of the figures. 
Away from this range where GIM mechanism is severe, 
${\cal B}(b^\prime\to cW)$ decreases as we increase $m_{b^\prime}$ 
while ${\cal B}(b^\prime\to b Z)$ and ${\cal B}(b^\prime\to bH)$ grow. 
It appears from this plot that for $m_{b^\prime}> 150$ GeV,
the three decays we are considering have 
the same order of magnitude with no single mode fully dominating.
For example, for $r_{\rm CKM} \simeq 0.004$
and $m_{b^\prime}$, $m_H =$ 170, 115 GeV,
the modes $cW : bZ : bH \simeq 2 : 2 : 1$ in rate.

%%%%%%%%%%%%
\begin{figure}[b!]
%\vskip.2cm
\centerline{{\epsfxsize2.1 in \epsffile{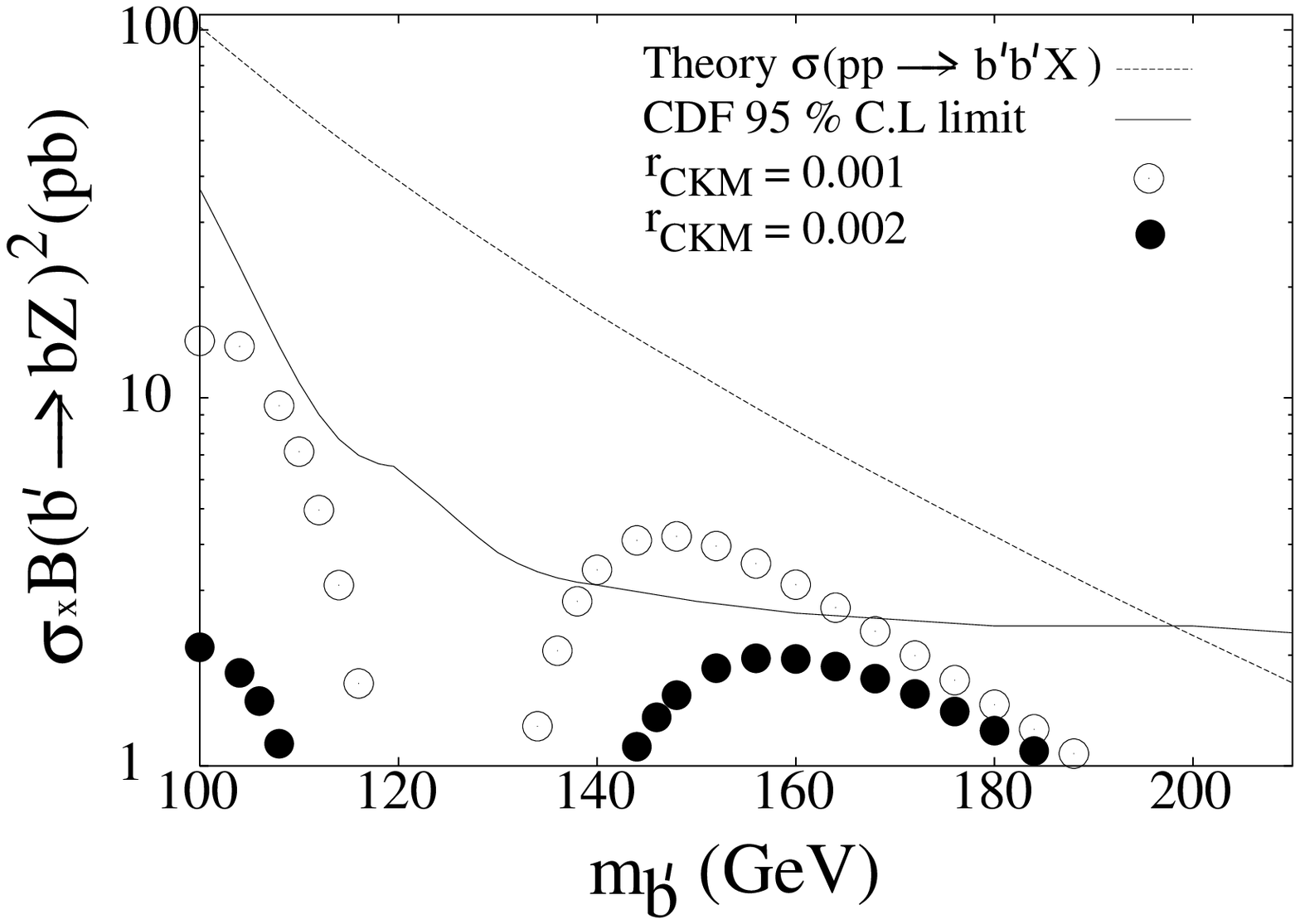}}}
%\smallskip
\vskip.8cm
\caption{
Comparison of CDF search for $b^\prime \to bZ$
and our scenario with $m_H = 115$, $\Delta_Q = 50$ GeV.
The dotted curve is the predicted 
$\sigma(p\bar{p}\to b^\prime \bar{ b^\prime})$ at 1.8 TeV, 
while the solid curve is the 95\% CL upper limit on 
$\sigma(p\bar{p}\to b^\prime \bar{b^\prime})
 \times [{\cal B}(b^\prime \to bZ)]^2$ [10].
The open (black) circles
correspond to ${\cal B}(b^\prime\to bZ)$
as computed in our model for $r_{\rm CKM}=0.001$ ($0.002$).
}
\end{figure}
%%%%%%%%%%%%

%We see that the two FCNC decays $b^\prime\to b Z$ and $b^\prime\to bH$ 
%and the CC decay $b^\prime\to c W$ can be of the same order of
%magnitude in a naturally preferred range of parameter space. 
%
To illustrate that this scenario can evade CDF 
search for $b^\prime \to bZ$,
we reproduce Fig. 2 of \cite{CDF} in our Fig. 4.
The dotted curve is the predicted cross section
$\sigma(p\bar{p}\to b^\prime \bar{ b^\prime})$ at 1.8 TeV, 
the solid curve corresponds to the 95\% CL upper limit on 
$\sigma(p\bar{p}\to b^\prime \bar{b^\prime})
 \times [{\cal B}(b^\prime \to bZ)]^2$.
From the crossing of the two curves, 
CDF rules out $m_{b^\prime} \lesssim 200$ GeV 
if ${\cal B}(b^\prime \to bZ) = 100\%$.
Our results are shown as open and black circles
for $r_{\rm CKM} =$ 0.001 and 0.002, respectively.
For larger $r_{\rm CKM}$ values, they drop out from the plot.
We have held the splitting 
$\Delta_Q = m_{t^\prime} - m_{b^\prime} =50$ GeV fixed.
This leads to the valley around $m_{b^\prime} \sim 125$ GeV,
caused by $m_{t^\prime} \simeq m_t$.
Very low cross section in $b^\prime\to bZ$ mode
can evade the search of \cite{CDF},
but it would likely be ruled out by past top search
since $b^\prime \to cW$ is predominant \cite{D0}. 
The reverse situation illustrates the power of
the CDF study: the region 
GeV $140 < m_{b^\prime} <160$ GeV for $r_{\rm CKM}=0.001$
is ruled out because ${\cal B}(b^\prime\to bZ)$ is predominant.
However, it is clear that 
the theoretical prediction is smaller than
the 95\% CL upper limit of CDF for a broad range
of parameter space,
as illustrated by the $r_{\rm CKM}=0.002$
case for $m_{b^\prime} \gtrsim 140$ GeV.
This is consistent with Figs. 2 and 3.
We note that a light $m_{b^\prime}$ around 110 GeV
is still allowed, although this case would not
help in Higgs search.

\begin{table}[t!]
\caption{Comparison of $\sigma (p\bar{p} \to b^\prime\bar{b^\prime}
\to WH\bar{c}b$, $ZH\bar{b}b)$ and $\sigma (p\bar{p}\to WH$, $ZH)$ (in pb)
for $m_{b^\prime}$, $m_{H}$, $\Delta_Q$ (in GeV) and 
$r_{\rm CKM} \equiv \vert {V_{cb^\prime}}/{V_{tb}V_{tb^\prime}} \vert$
taken from Fig. 2.  
\label{tabone} 
}
\begin{tabular}{c|lc|lc}
$\;m_{b^\prime}$, \ $m_{H}$, $\; \Delta_Q$; \ $r_{\rm CKM}$
 & $WH$  &  $WH\bar{c}b$ &   $ZH$  & $ZH\bar{b}b$  \\ \hline
$130$, \ $\;115$, \ $\;20$; \ \ $0.001$  &
  0.20  &  0.65  &  0.11  & 1.49 \\ 
$130$, \ $\;115$, \ $\;20$; \ \ $0.002$  &
  0.20  &  0.78  &  0.11 &  0.45 \\ 
$130$, \ $\;115$, \ $\;20$; \ \ $0.004$  &
  0.20  &  0.39  & 0.11  &  0.06 \\ \hline
$160$, \ $\;130$, \ $\;40$; \ \ $0.001$  &
  0.13  &  0.60  &  0.07  &  2.14 \\
$160$, \ $\;130$, \ $\;40$;  \ \ $0.002$  &
  0.13  &  1.00  &  0.07  &  0.88 \\
$160$, \ $\;130$, \ $\;40$; \ \ $0.004$  &
  0.13  &  0.68  &  0.07  &  0.15
\end{tabular}
\end{table}

We now compare the cross section for 
$p\bar{p}\to b^\prime\bar{b^\prime}\to (\bar{c}W)(bH)$ or $(\bar{b}Z)(bH)$
with the direct Higgs production mechanism 
$p\bar{p}\to WH$, $ZH$. 
For illustration and clarity, we recapitulate our numerical results from
Fig.~2 for Run II energies (2 TeV) at Tevatron in Table~I.
The cross sections for $p\bar{p}\to WH,\ ZH$
 and $p\bar{p}\to b^\prime\bar{b^\prime}$  
are taken from \cite{spira} and \cite{bb}, respectively. 
It is clear that 
the cross sections for $WH\bar{c}b$ or $ZH\bar{b}b$ are 
larger than the corresponding ones for 
direct associated $WH$ or $ZH$ production,
unless 
$r_{\rm CKM} = \vert V_{cb^\prime}/{V_{tb}V_{tb^\prime}}\vert$ 
becomes considerably larger than few $\times 10^{-3}$,
or when $b^\prime \to bH$ is kinematically suppressed.
Thus, our suggestion should be welcome news for
the Higgs search program at Tevatron Run II.
In any case, the search for $b^\prime$ below the top
should continue by taking into account the $b^\prime\to cW$ mode.
By so doing, one may uncover the Higgs boson!

Curiously, there are some indications that
the ``cocktail solution" of $b^\prime\to cW$, $bZ$ and $bH$
should be taken seriously and hence revisited even for Run I data.
It is known that
the $t\bar t$ events at the Tevatron have some irregularities
that are, though not yet statistically significant,
somewhat tantalizing.
First,
{\it both} CDF and D0 \cite{dilepton}
report a lower $m_t \simeq 167$--$168$~GeV
in the dilepton channel,
where $b$-tagging is not used.
Second, 
for single lepton plus jets channel,
D0 and CDF \cite{lj} are in good agreement on $m_t$,
but the CDF cross section extracted from soft lepton tag (SLT)
is almost twice as high from the displaced vertex (SVX) tag,
with fitted $m_t$ as low as 142 GeV!
Third,
the all hadronic study of CDF \cite{Had},
which relies heavily on $b$-tagging, 
gives $m_t \simeq 186$~GeV, the highest of all studies.
Interestingly,
if one demands two SVX-tagged $b$-jets
for single lepton plus jets sample,
the fit following CDF dilepton procedure 
also gives a high mass of $m_t \simeq 182$ GeV \cite{dilepton}.

For sake of illustration,
we show that the combination $m_t \sim 175$ GeV
and $m_{b^\prime},\ m_{t^\prime} \sim 160$, 210 GeV, 
with $t,\ t^\prime\to bW$ \cite{tp}
and the ``cocktail solution" of $b^\prime\to cW \sim bZ > bH$,
can account for these curiosities.
For dileptons without $b$-tagging, one largely probes 
$b\bar bW^+W^-$ for top and $t^\prime$, 
and $c\bar c W^+W^-$ for $b^\prime$. 
One would get lower ``$m_t$" and a somewhat larger cross section.
For single lepton plus jets with SVX $b$-tag, 
one is less sensitive to $b^\prime\bar b^\prime$,
thereby getting an average ``$m_t$".
However, applying SLT tag but no SVX tag,
one is then sensitive to both $b$ and $c$ semileptonic decays
with similar efficiencies,
and one would be more sensitive
to $b^\prime \to cW$ decay which has 
larger $b^\prime\bar b^\prime$ cross section
and a lower fitted ``$m_t$".
For all hadronic final state,
since one demands SVX $b$-tag to suppress QCD background,
one is more sensitive to $t,\ t^\prime \to bW$
hence a higher ``$m_t$" is found.

The pattern in cross sections could also be reflecting
the presence of both $b^\prime$ and $t^\prime$ besides the top,
as already stated in the larger SLT vs SVX tagged cross section.
The dilepton and all hadronic cross sections are also 
somewhat larger than theory expectation of order 5 pb$^{-1}$.
However, not much more can be said because of 
experimental errors at the Run I level of statistics.
One would also need detailed knowledge of experimental efficiencies.
Although one cannot draw a definite conclusion,
it may still be worthwhile even
to reinvestigate the Run I ``$t\bar t$" data,
keeping in mind the possibility of the ``cocktail solution":
There may actually be charm jet content in $t\bar t$-like events.
At Run~II, such a study would be imperative,
for not only one would have the statistical power to distinguish,
a more exciting Higgs search program could be at stake!
One could be discovering the Higgs together with two new quarks.

We have shown in this analysis that, 
once the constraints from precision measurements
are taken into account, 
$m_{b^\prime}\sim m_{t^\prime}\sim m_t$
could be the case.
This suppresses FCNC $b^\prime\to bZ$, $bH$ decays 
so the CC $b^\prime\to cW$ decay becomes important. 
It should be welcome since this is just what is needed to
evade the CDF null search for $b^\prime\to bZ$.
In a rather plausible parameter space and 
if the GIM suppression of FCNC modes is not overly strict,
one can have the ``cocktail solution" of
$b^\prime\to cW \sim bZ > bH$,
and there could be actual charm jet content of
observed $t\bar t$ events.
Such a signal should be considered
for better scrutiny of top-like events,
and might uncover the Higgs boson handily at Tevatron Run II
via $p\bar p\to \bar{c}bWH +X$ or $b\bar{b}ZH +X$ signatures.

A.A. is on leave of absence from Department of Mathematics, 
FSTT, P.O.B416 Tangier, Morocco.
This work is supported in part by
NSC-89-2112-M-002-063 and 0129, 
and the MOE CosPA Project. 
We thank Jaroslav Antos and Paoti Chang for discussions.


\begin{thebibliography}{99}
\bibitem{tW*} We take into account $b^\prime \rightarrow t W^*$ 
for $175 <m_{b^\prime} <200$ GeV, but
it is numerically insignificant.
\bibitem{tp} In our scenario or small $t^\prime$--$t$ splitting
and large $V_{t^\prime b}$, 
the $t^\prime$ quark also dominantly decays via $bW$ mode.
\end{thebibliography}
\end{document}